\documentclass[page-classic]{epl2} 
\usepackage{graphicx}
\usepackage{dcolumn}
\usepackage{bm}
\usepackage{color}
\usepackage[pdftitle={Prepared for EPL},bookmarks=true,citecolor=red,colorlinks=false]{hyperref}
\usepackage{amsmath}
\usepackage{amssymb}
\usepackage{amsthm}
\usepackage{wasysym}
\usepackage[displaymath]{lineno}
 \linenumbers
\usepackage{ulem}

\newcommand{\figref}[1]{~\ref{#1}}
\renewcommand{\eqref}[1]{~(\ref{#1})}
\title{Autocorrelation function of velocity increments  {time series} in fully developed turbulence}

\author{Y.X. Huang\inst{1,2}\thanks{E-mail: \email{yongxianghuang@gmail.com}} \and F. G. Schmitt\inst{2}\thanks{E-mail: \email{francois.schmitt@univ-lille1.fr}} \and Z.M. Lu\inst{1} \and Y.L. Liu\inst{1}}
\shortauthor{Y.X. Huang \etal}

\institute{
 \inst{1} Shanghai Institute of Applied Mathematics and Mechanics, Shanghai University, 200072 Shanghai,  China\\
  \inst{2} Universit\'e des Sciences et Technologies de Lille - Lille 1, CNRS, Laboratory of Oceanology and Geosciences,
  UMR 8187 LOG,  62930 Wimereux, France\\

}

\abstract{
In fully developed turbulence, the velocity field possesses long-range correlations,
denoted by a scaling power spectrum or structure functions. Here we consider the autocorrelation function of velocity increment $ {\Delta u_{\ell}(t)}$ at separation  { time} $\ell$. Anselmet et al. [Anselmet et al. J. Fluid Mech. \textbf{140}, 63 (1984)] have found
that the  autocorrelation function of velocity increment has a minimum value, whose location  is  approximately equal to $\ell$. Taking  statistical stationary assumption, we link the velocity
increment and the autocorrelation function with the power spectrum of the
original variable. We then propose an
 analytical model of the autocorrelation function.  With this model, we prove that the
location of the minimum autocorrelation function is exactly equal to the separation  {time} $\ell$ when the scaling of
the power spectrum of the original variable belongs to the range $0<\beta<2$.  This
model also suggests a power law expression for the minimum autocorrelation. Considering
the cumulative function of  the autocorrelation function, it is shown that the main contribution to the autocorrelation function comes from the large scale part. Finally we argue that the autocorrelation function is a better
indicator of the inertial range  than the second order structure function. }

\begin{document}

\maketitle
\section{Introduction}

Turbulence is characterized by power law of the velocity spectrum \cite{Kolmogorov1941} and
structure functions in the inertial range \cite{Monin1971,frisch1995}.  This is associated to  long-range power-law correlations for the dissipation or absolute  value of the
velocity increment. Here we consider the autocorrelation of velocity increments (without
absolute value), inspired by a remark found in Anselmet et al. (1984) \cite{Anselmet1984}. In this reference, it is  found that the location of the minimum value of the
autocorrelation function $\Gamma(\tau)$ of velocity increment { $\Delta u_{\ell}(t)$,
defined as 
\begin{equation}
\Delta u_{\ell}(t)=u(t+\ell)-u(t)
\end{equation}
}
 of fully
developed turbulence with  { time} separation $\ell$ is approximately equal to $\ell$. The autocorrelation  function of  {this time series} is defined as
\begin{equation}
  \Gamma(\tau)={\langle (V_{\ell}(t)-\mu)(V_{\ell}(t-\tau)-\mu)\rangle}
\end{equation}
where  {$V_{\ell}(t)=\Delta u_{\ell}(t)$}, $\mu$ is the mean value of  {$V$}, and $\tau>0$ is the time lag.

This paper mainly presents analytical results. In first section we present the database considered  here as an illustration of the property which is studied. The next section presents theoretical studies. The last section provides a discussion.

\begin{figure}[htb]
\centering
 \includegraphics[width=0.95\linewidth]{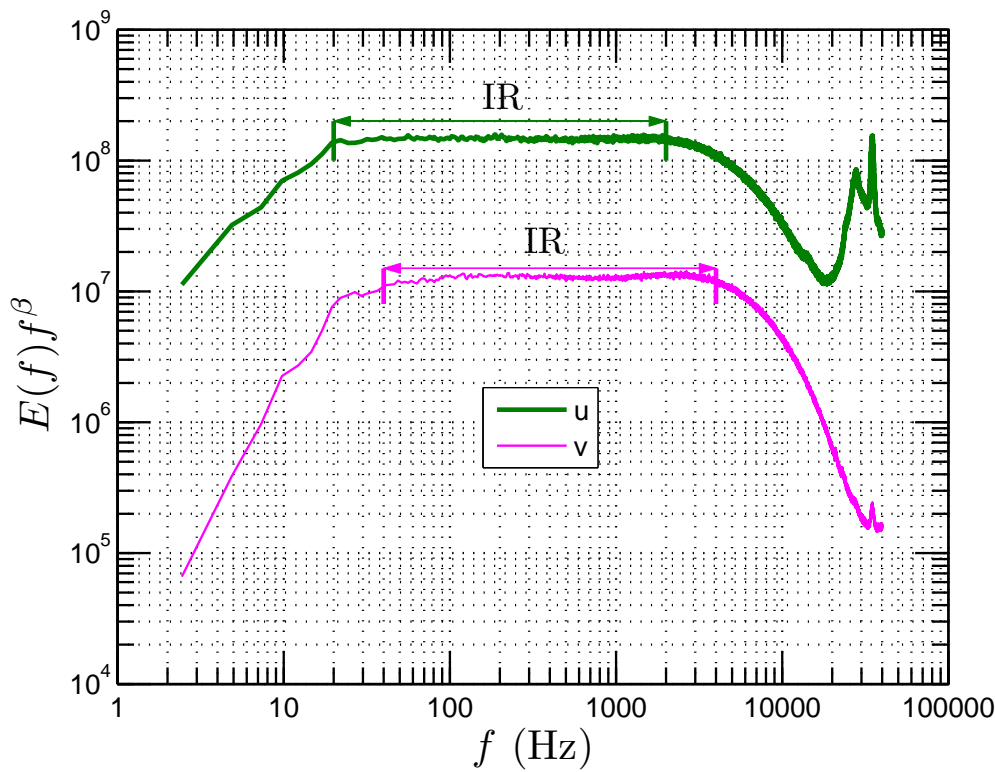}
  \caption{Compensated spectrum $E(f)f^{\beta}$ of  {
  streamwise (longitudinal) ($\beta\simeq1.63$)} and  { spanwise
  (transverse)($\beta\simeq1.62$)}  velocity,   where $\beta$ is the corresponding power law estimated from
the power spectrum. The plateau is observed
on the range $20<f<2000$ Hz and $40<f<4000$ Hz for  {
  streamwise (longitudinal)} and  { spanwise (transverse)}
velocity, respectively. }\label{fig:compsp}
\end{figure}

\begin{figure}[htb]
\centering
 \includegraphics[width=0.95\linewidth]{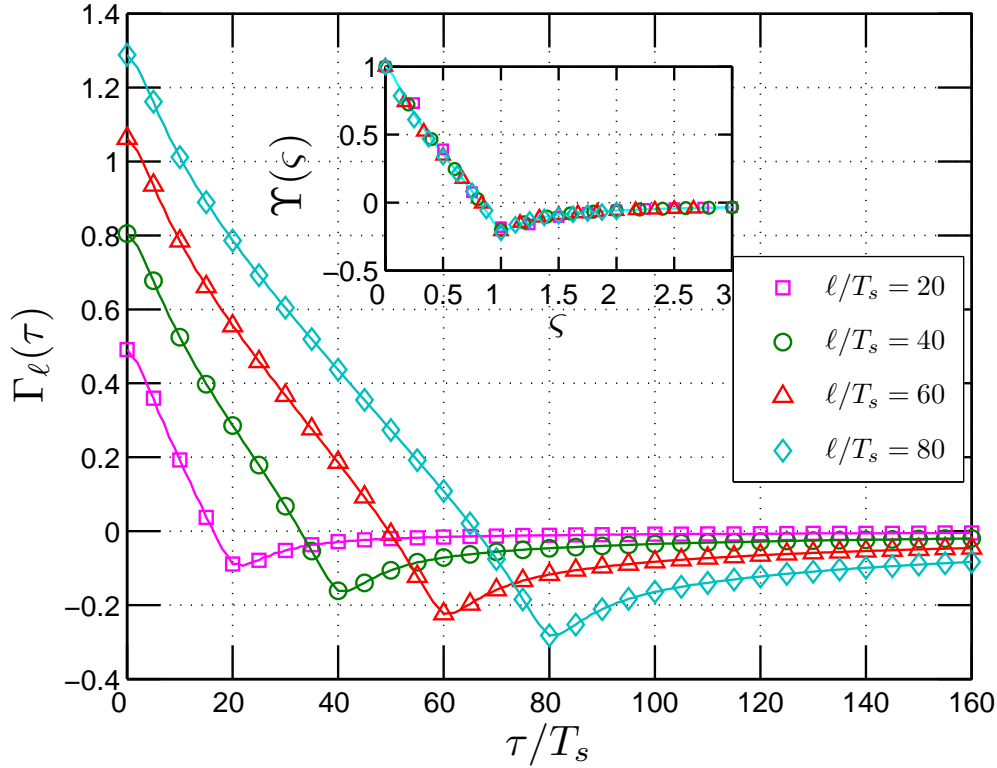}
  \caption{Autocorrelation function $\Gamma_{\ell}(\tau)$ of the velocity increment  {$\Delta u_\ell(t)$}
   estimated from an experimental homogeneous and nearly isotropy turbulence time series
   with various increments $\ell$.  The location of  the minimum value is very close to the separation  {time}
 $\ell$. The inset shows the rescaled autocorrelation
function $\Upsilon(\varsigma)$.}\label{fig:acf}
\end{figure}

\begin{figure}[htb]
\centering
 \includegraphics[width=0.95\linewidth]{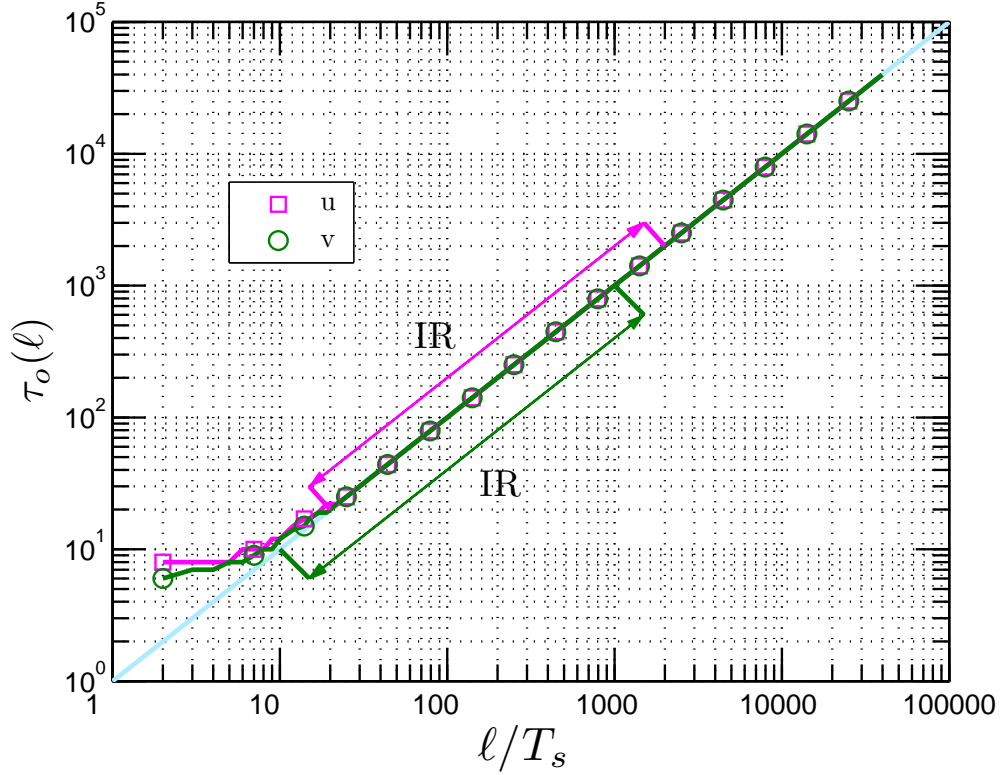}
  \caption{Location $\tau_{o}(\ell)$ of the minimum value of the autocorrelation
  function estimated from experimental data, where the inertial range is
marked as IR.  The solid line indicates $\tau_{o}(\ell)=\ell$.}\label{fig:tauo}
\end{figure}

\begin{figure}[htb]
\centering
 \includegraphics[width=0.95\linewidth]{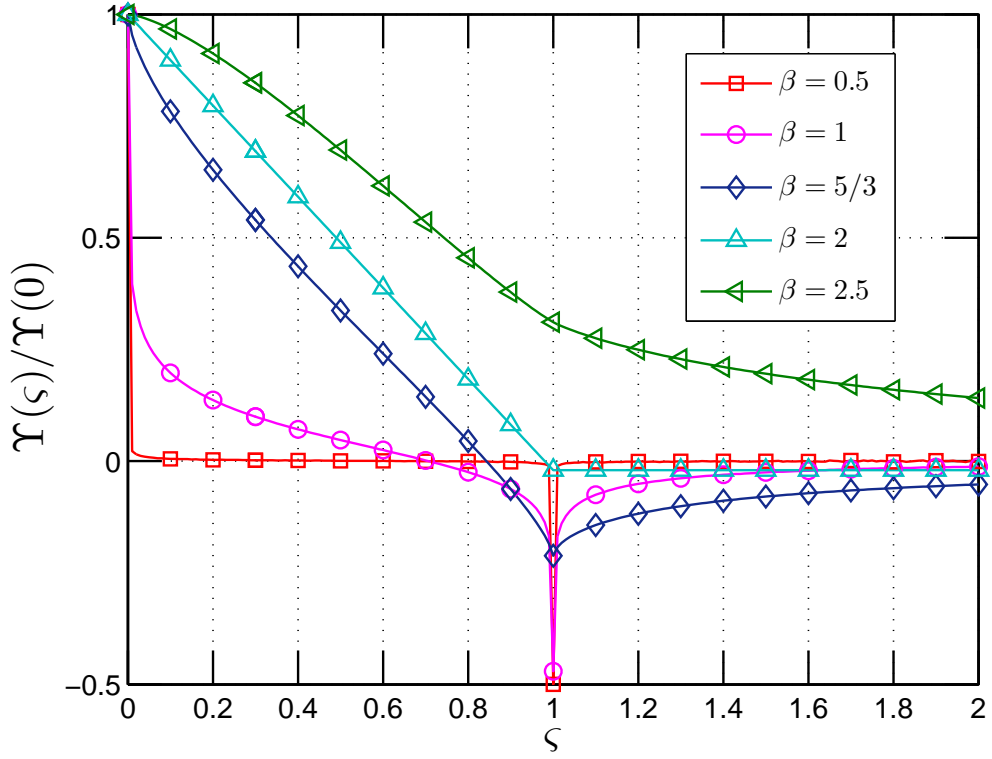}
  \caption{Numerical solution of the rescaled autocorrelation function  $\Upsilon(\varsigma)$
  with various $\beta$ from 0.5 to 2.5 estimated from eq.\eqref{eq:rhop}.  }\label{fig:nrho}
\end{figure}

\begin{figure}[!htb]
\centering
 \includegraphics[width=0.95\linewidth]{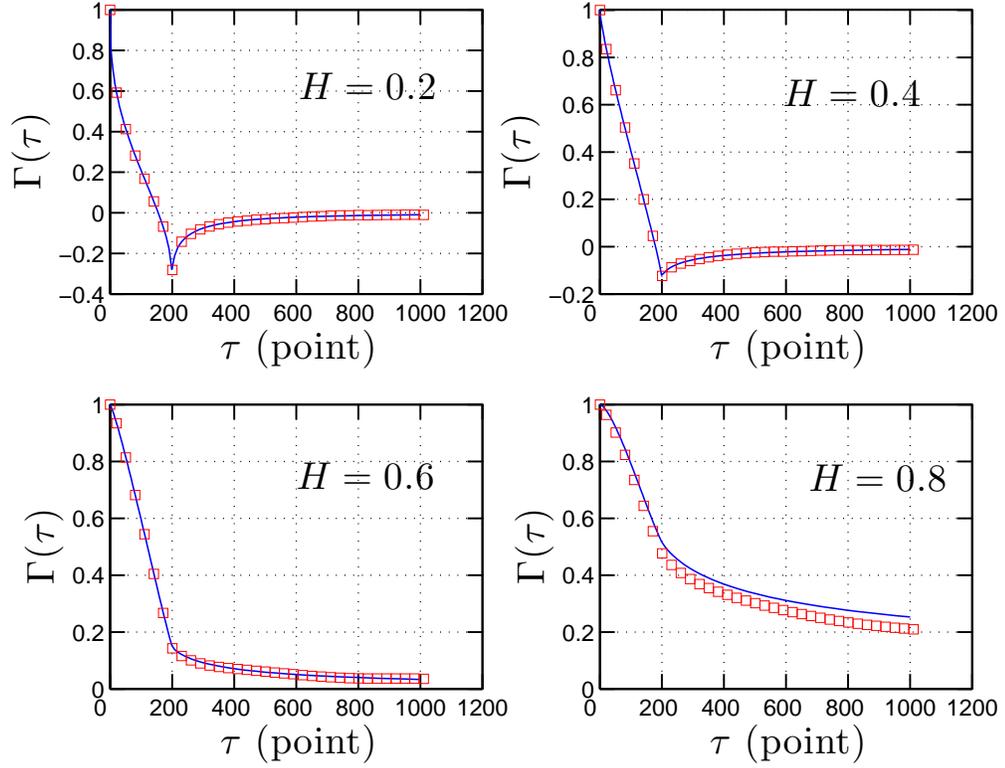}
  \caption{ {Comparison of the autocorrelation function, which is predicted
  by eq.\eqref{eq:correlation} (solid line) and estimated from fBm
simulation ($\square$) with $\ell=200$ points.}}\label{fig:fbm}
\end{figure}

\begin{figure}[htb]
\centering
 \includegraphics[width=0.95\linewidth]{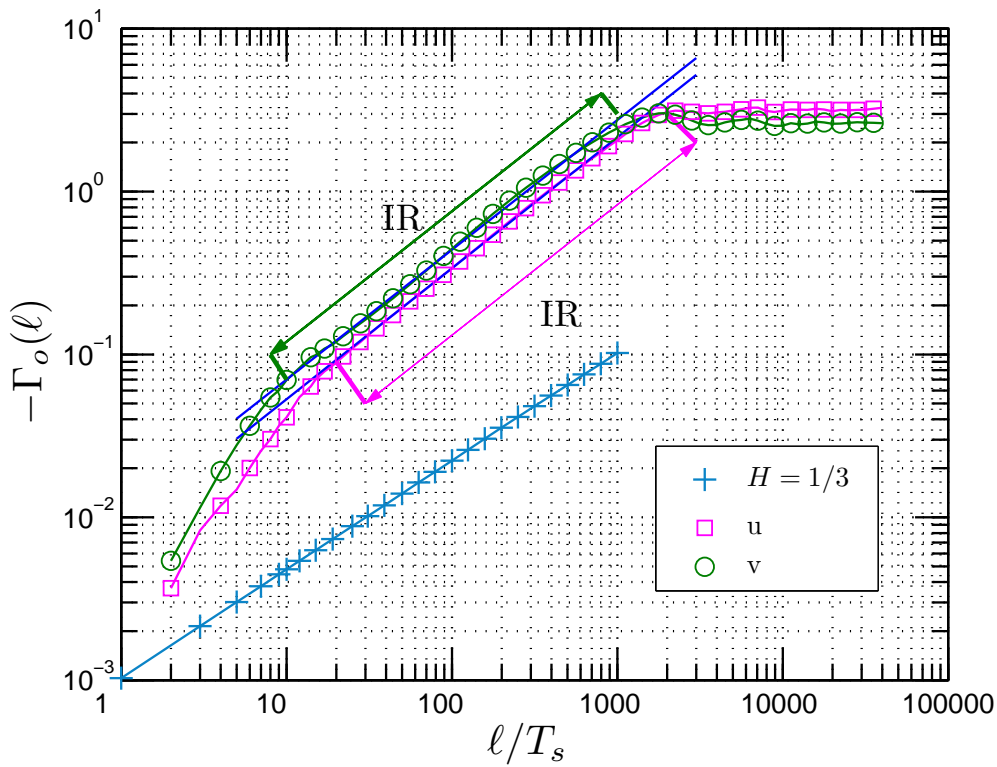}
  \caption{Representation of the minima value $\Gamma_{o}(\ell)$ of the autocorrelation
function estimated from synthesized fBm time series with $H=1/3$ ($+$), and the experimental data for  { streamwise (longitudinal)} ($\square$) and  { spanwise (transverse)} ($\ocircle$) turbulent velocity components, where the corresponding inertial range is denoted as IR.
  Power law behaviour is observed  with scaling exponent $\beta-1=2/3$ and $\beta-1=0.78\pm 0.04$ for fBm and turbulent velocity, respectively.   }\label{fig:beta}
\end{figure}

\begin{figure}[htb]
\centering
 \includegraphics[width=0.95\linewidth]{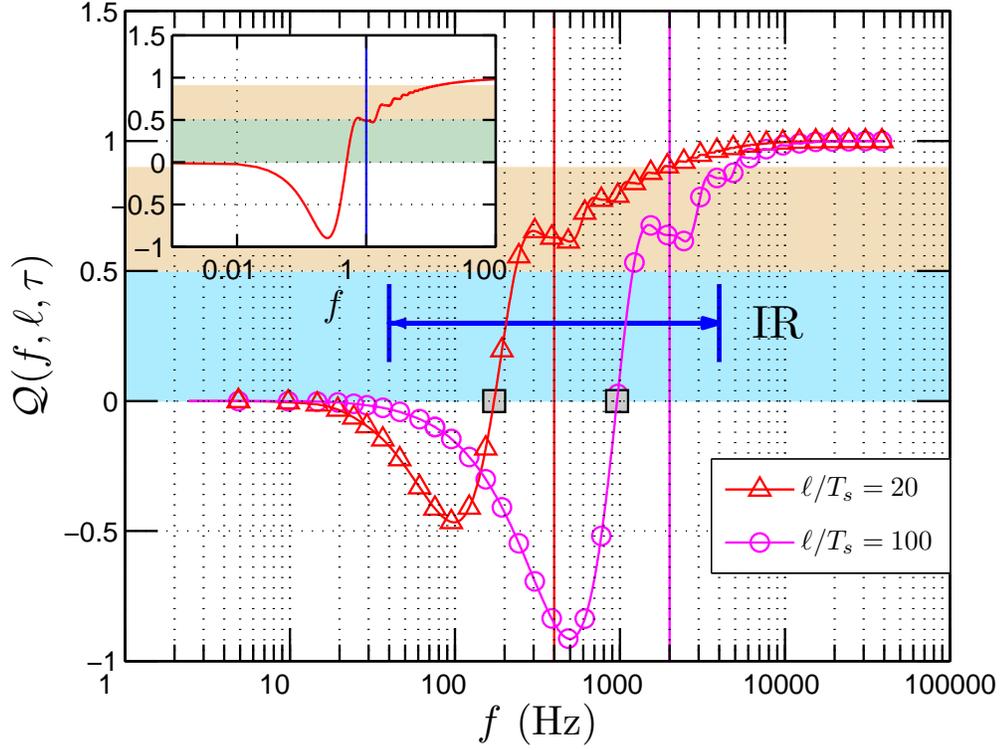}
  \caption{Cumulative  function $\mathcal{Q}(f,\ell,\tau)$ estimated from turbulent
  experimental data for   { spanwise (transverse)}  velocity with  $\tau=\ell$ in the inertial range, where the numerical solution is shown as inset with $\ell=1$.
The inertial range is denoted as IR.  Vertical solid lines demonstrate the corresponding scale in spectral space. }\label{fig:cumulant}
\end{figure}

\begin{figure}[htb]
\centering
 \includegraphics[width=0.95\linewidth]{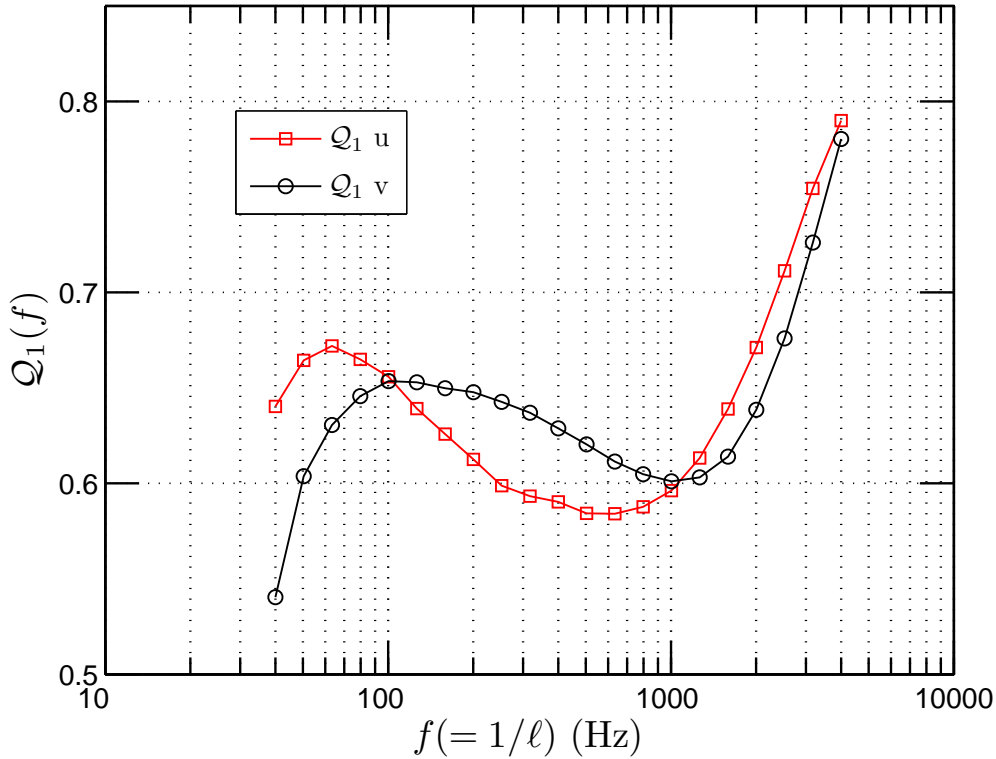}
  \caption{Cumulative function $\mathcal{Q}_1(f)$ estimated from turbulent
  experimental data for both  { streamwise (longitudinal)} and  { spanwise (transverse)} velocity with various $\ell$. The numerical solution is $\mathcal{Q}_1\simeq 0.5$.}\label{fig:Q1}
\end{figure}

\begin{figure}[htb]
\centering
 \includegraphics[width=0.95\linewidth]{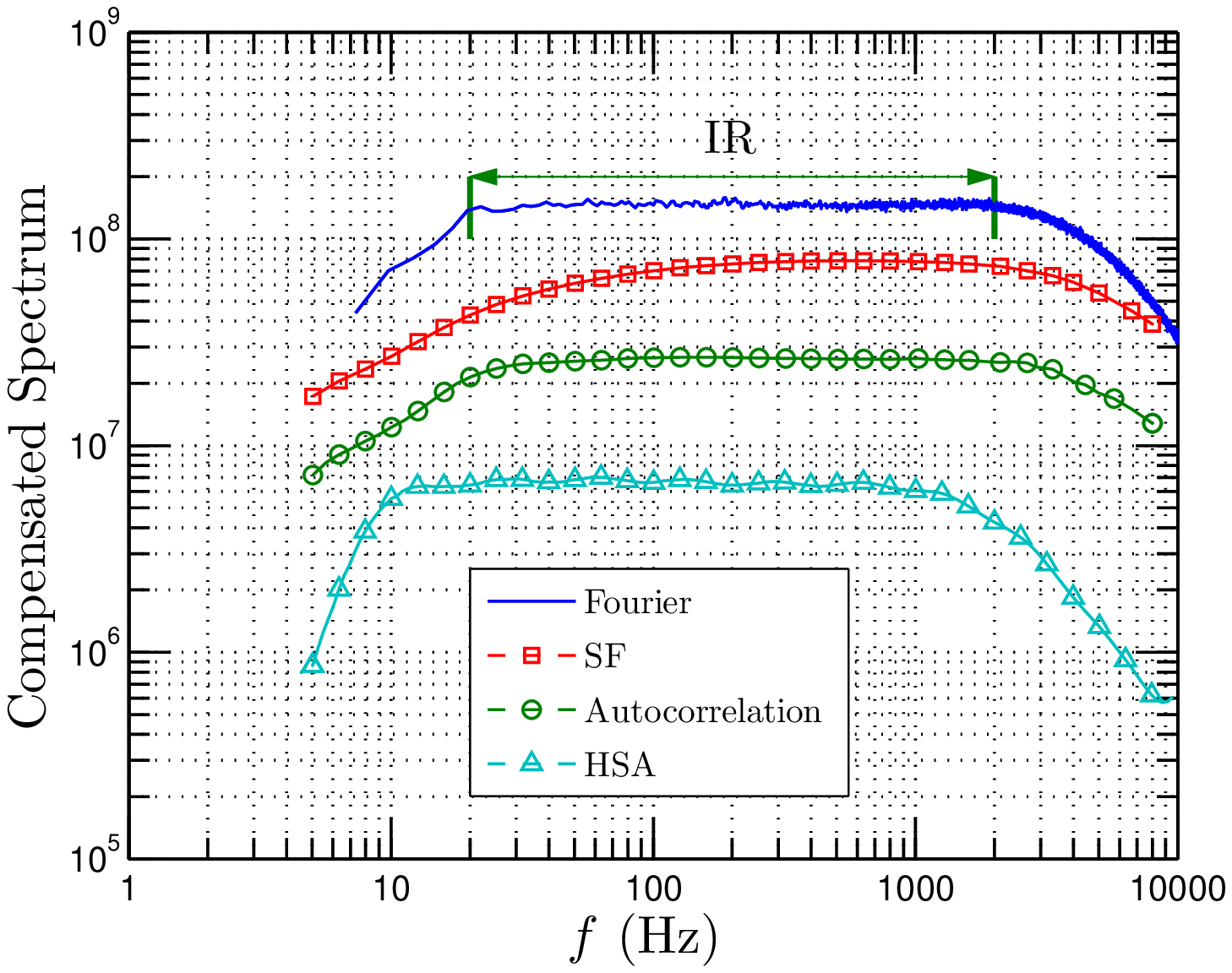}
  \caption{Comparison of the inertial range for the  {streamwise (longitudinal)} velocity. They are estimated  {directly} by the Fourier power spectrum, the second order structure function,the Hilbert spectral analysis and the autocorrelation function.}\label{fig:CP}
\end{figure}

\section{Experimental analysis of the  autocorrelation function of velocity increments}
We consider here a turbulence velocity time series
obtained from an experimental homogeneous and nearly isotropic turbulent flow
at downstream $x/M=20$, where $M$ is the mesh size. The flow is characterized by the Taylor microscale based Reynolds number
$Re_{\lambda}=720$\cite{kang2003dta}.  {The sampling frequency is
$f_s=40\,$kHz and a low-pass filter at a frequency $20\,$kHz is applied on
the experimental data. The sampling time is $30\,$s, and the  number
of data points per channel for each measurement is $1.2\times10^6$. We have
120 realizations with four channels. The total number of data points at this
location is $5.76\times
10^8$. The mean velocity is $12\,\mathrm{ms^{-1}}$.  The rms velocity is
1.85 and $1.64\,\mathrm{ms^{-1}}$ for streamwise (longitudinal) and spanwise
(transverse) velocity component. The Kolmogorov scale $\eta$ and the Taylor
microscale $\lambda$ are 0.11\,mm and 5.84\,mm respectively. Let us note
here $T_s=1/f_s$ time resolution of these measurements.}
This  data demonstrates an  inertial range over two decades \cite{kang2003dta},
see  {a compensated spectrum $E(f)f^{\beta}$ in fig.\figref{fig:compsp},
where $\beta\simeq1.63$ and $\beta\simeq1.62$ for streamwise (longitudinal) and spanwise (transverse)
velocity respectively}.
We show  the autocorrelation function $\Gamma_{\ell}(\tau)$
directly estimated from these data in  fig.\figref{fig:acf}.   Graphically, the location $\tau_o$ of the minimum value of each curve is very
close to $\ell$, which confirms Anselmet's observation \cite{Anselmet1984}.
Let us define
\begin{equation}
\Gamma_{o}(\ell)=\min_{\tau}\{ \Gamma_{\ell}(\tau)\}
\end{equation}
and $\tau_{o}$ the location of the minimum value
\begin{equation}
\Gamma_{o}(\ell)=\Gamma_{\ell}(\tau_{o}(\ell))
\end{equation}
 We show the estimated $\tau_o(\ell)$ on the range  {$2<\ell/T_s<40000$}   in
 fig.\figref{fig:tauo}, where the inertial range is indicated by IR.  It shows that when $\ell$ is greater than  {$20T_s$}, $\tau_o$ is
very close to $\ell$ even when $\ell$ is in the forcing range, in agreement with the remark of Anselmet et al. \cite{Anselmet1984}. In the following, we show this analytically.

\section{Autocorrelation function}
Considering the statistical stationary assumption\cite{frisch1995}, we represent  {$u(t)$}
in Fourier space, which is written as
 {\begin{equation}
\hat{U}(f)=\mathcal{F}(u(t))=\int_{-\infty}^{+\infty} u(t) e^{ -2\pi i f t} \upd t
\end{equation}}
 where $\mathcal{F}$ means Fourier transform and $f$ is the frequency.
 Thus, the Fourier transform of the velocity increment  {$\Delta u_\ell(t)$} is written as
\begin{equation}
 {S}_{\ell}(f)= \mathcal{F}( {\Delta u_\ell(t)})=\hat{U}(f)(e^{2\pi i f \ell}-1)
\end{equation}
where  {$\Delta u_\ell(t)=u(t+\ell)-u(t)$}. Hence, the 1D power spectral density function of velocity increments $E_{\Delta}(f)$ is  expressed as
\begin{equation}
  E_{\Delta}(f)=\vert S_{\ell}(f)\vert^2=E_{v}(f)(1-\cos(2\pi f \ell))\label{eq:relation}
\end{equation}
where $E_{v}(f)=2\vert \hat{U}(f)\vert^2$ is the velocity power spectrum \cite{frisch1995}.
It is  clear that the velocity increment operator acts a kind of filter, where the frequencies  $f_{\Delta}=n/\ell$, $n=0, 1, 2\cdots$, are filtered.

Let us consider now the autocorrelation function of the  increment.
The Wiener-Khinchin theorem relates the autocorrelation function to the power spectral density via the Fourier transform\cite{Percival1993,frisch1995}
  \begin{equation}
    \Gamma_{\ell}(\tau)=\int_{0}^{+\infty} E_{\Delta}(f) \cos(2\pi f \tau) \upd f\label{eq:acf}
  \end{equation}
   {The theorem can be applied to wide-sense-stationary random processes, signals whose Fourier transforms may  not exist, using the definition of autocorrelation function in terms of expected value rather than an infinite integral\cite{Percival1993}.}
Substituting eq.\eqref{eq:relation} into the above equation, and assuming a power law for the spectrum { (a hypothesis of similarity)}
  \begin{equation}
  E_{v}(f)=c f^{-\beta},\quad c>0
  \end{equation}
we obtain
  \begin{equation}
      \Gamma_{\ell}(\tau)=c\int_{0}^{+\infty} f^{-\beta}(1-\cos(2\pi f \ell)) \cos(2\pi f \tau)\upd f\label{eq:rhop}
 \end{equation}
The convergence condition requires $0<\beta<3$. It implies a rescaled relation, using scaling transformation inside the integral.
This can be estimated  by taking $\ell'=\lambda \ell$, $f'= f \lambda$, $\tau'=\tau/\lambda$ for $\lambda>0$, providing the identity
\begin{equation}
 \Gamma_{\lambda\ell}(\tau)= \Gamma_{\ell}(\tau/\lambda)\lambda^{\beta-1}\label{eq:power1}
\end{equation}
If we take $\ell=1$ and replace $\lambda$ by $\ell$, we then have
\begin{equation}
 \Gamma_{\ell}(\tau)= \Gamma_{1}(\tau/\ell)\ell^{\beta-1}\label{eq:power}
\end{equation}
Thus, we have a universal  autocorrelation function
\begin{equation}
{\Gamma_{\ell}(\ell \varsigma)}{\ell^{1-\beta}}
=\Upsilon(\varsigma)=\Gamma_{1}(\varsigma)
\end{equation}
This rescaled universal autocorrelation function is shown as inset in fig.\figref{fig:acf}.
 A derivative of eq.\eqref{eq:power1} gives $\Gamma'_{\lambda\ell}(\tau)=\Gamma'_{\ell}(\tau/\lambda)\lambda^{\beta-2}$. The minimum value of the left-hand side is $\tau=\tau_{o}(\lambda \ell)$, verifying  $\Gamma'_{\lambda\ell}(\tau_{o}(\lambda\ell))=0$ and for this value we have also $\Gamma'_{\ell}(\tau_{o}(\lambda\ell)/\lambda)=0$. This shows that $\tau_{o}(\ell)=\tau_{o}(\lambda \ell)/\lambda$.
Taking again $\ell=1$ and $\lambda=\ell$, we have
\begin{equation}
  \tau_{o}(\ell)=\ell\tau_{o}(1)\label{eq:tau}
\end{equation}
Showing that $\tau_{o}(\ell)$ is proportional to $\ell$ in the scaling range
 {(when $\ell$ belongs to the inertial range)}. With the definition of $\Gamma_{o}(\ell)=\Gamma_{\ell}(\tau_{o}(\ell))$ we have, also using eq.\eqref{eq:power1}, for   $\tau=\tau_{o}(\lambda\ell)$:
\begin{equation}
\begin{array}{ll}
  \Gamma_{\lambda\ell}(\tau_{o}(\lambda\ell))&=\Gamma_{\ell}(\tau_{o}(\lambda\ell)/\lambda)\lambda^{\beta-1}\\
  &=\Gamma_{\ell}(\tau_{o}(\ell))\lambda^{\beta-1}
\end{array}
\end{equation}
Hence $\Gamma_{o}(\lambda \ell)=\lambda^{\beta-1}\Gamma_{o}(\ell)$ or
\begin{equation}
  \Gamma_{o}(\ell)=\Gamma_{o}(1)\ell^{\beta-1}
\end{equation}

We now consider the location $\tau_{o}(1)$ of the autocorrelation function for $\ell=1$. We take the first derivative of eq.\eqref{eq:rhop}, written for $\ell=1$
\begin{equation}
 \mathcal{P}(\tau)=\frac{\upd \Gamma_{1}(\tau)}{\upd \tau}=- \int_{0}^{+\infty} f^{1-\beta}(1-\cos(2\pi f ))\sin(2
 \pi f \tau)\upd f\label{eq:drho}
\end{equation}
where we left out the constant in the integral. The  same rescaling calculation leads to the following expression
\begin{equation}
\begin{array}{l}
\mathcal{P}(\tau)=\left[ (1+1/\tau)^{\beta-2}+(1-1/\tau)^{\beta-2}-2 \right]M/2, \tau \ne 1\\
\mathcal{P}(\tau)=\left(2^{\beta-3}-1 \right)M,\quad \quad  \tau=1
\end{array}
\label{eq:derivative}
\end{equation}
where  $M=\int_{0}^{+\infty} x^{1-\beta}(1-\cos(2\pi x ))\sin(2
 \pi x \tau)\upd x$ and $M>0$ \cite{Samorodnitsky1994}.
The convergence condition requires $1<\beta<4$.
{When $\beta<2$, one can find that both  left and right limits of $\mathcal{P}(1)$ are infinite, but the definition of $\mathcal{P}(1)$ in eq.\eqref{eq:drho} is finite. Thus $\tau=1$  is a second type discontinuity point of eq.\eqref{eq:drho} \cite{Malik1992}.}
  It is easy to show that
\begin{equation}
\left\{
\begin{array}{rl}
&\mathcal{P}(\tau)<0, \tau\le 1\\
&\mathcal{P}(\tau)>0,\tau>1
\end{array}
\right.
\end{equation}
It means that  $\mathcal{P}(\tau)$ changes its sign from negative to positive
when $\tau$ is increasing from $\tau<1$ to $\tau>1$. In other words the autocorrelation function will take its minimum value at the location where $\tau$ is exactly equal to $1$. We thus see that $\tau_{o}(1)=1$ and hence $\tau_{o}(\ell)=\ell$ (eq.\eqref{eq:tau}).

\section{Numerical validation}

There is  no analytical solution  for  eq.\eqref{eq:rhop}. It is then solved here by a proper numerical algorithm. We perform a fourth order accurate Simpson rule numerical integration of eq.\eqref{eq:rhop} on
range $10^{-4}<f<10^{4}$ with $\ell=1$ for various $\beta$  {with
step $\Delta f=10^{-6}$}. We show the rescaled
numerical
solutions $\Upsilon(\varsigma)$ for various $\beta$ values in fig.\figref{fig:nrho}.  Graphically, as
what we have proved above, the location
$\tau_o(1)$ of the minimum autocorrelation function is exactly equal to
$1$ when $0<\beta<2$.

 {
For the fBm, the autocorrelation function
of the increments is known to be the following \cite{Biagini2008}
\begin{equation}
\Gamma_{\ell}(\tau)=\frac{1}{2}\left\{ (\tau+\ell)^{2H}+\vert \tau-\ell \vert^{2H} -\tau^{2H} \right\}\label{eq:correlation}
\end{equation}
where $\tau\ge 0$. We compare the autocorrelation (coefficient) function  estimated from
fBm simulation ($\square$, see bellow) with eq.\eqref{eq:correlation} (solid line) in fig.\figref{fig:fbm}, where
$\ell=200$ points. Graphically, eq.\eqref{eq:correlation} provides a very
good prediction with numerical simulation.  Based on this model, it is not
difficult to find that  $\Gamma_{o}(\ell)\sim\ell^{2H}$ when $0< H <1$, corresponding
to $1<\beta<3$, and $\tau_{o}(\ell)=\ell$
when $0<H<0.5$, corresponding to $1<\beta<2$. One can find that the validation
range of scaling
exponent $\beta$ is only a subset of Wiener-Khinchin theorem.}

We then check the power law for the minimum value of the autocorrelation function given in
eq.\eqref{eq:power}. We simulate 100 segments
of fractional Brownian
motion with length $10^6$ data points each,  by performing a Wavelet
based algorithm \cite{Abry1996}. We take db2 wavelet with $H=1/3$ (corresponding to the Hurst number of turbulent velocity). 
We plot the
estimated minima value $\Gamma_{o}(\ell)$ ($+$) of the autocorrelation function in
fig.\figref{fig:beta}. A power law behaviour is observed with the scaling exponent
$\beta-1=2/3$ as expected.
It confirms  eq.\eqref{eq:power} for fBm. We also plot  $\Gamma_{o}(\ell)$
estimated from turbulent experimental data for both  {streamwise
(longitudinal)} ($\square$) and  { spanwise (transverse)} ($\ocircle$)
velocity components in fig.\figref{fig:beta},
where the inertial range is marked by IR.
 Power law is observed on the corresponding inertial range   {with
 scaling exponent $\beta-1=0.78\pm0.04$. Due to the intermittency, this scaling exponent is larger than
2/3. The exact relation between this scaling exponent with intermittent
parameter should be investigated further in future.} The power law range
is almost the same as the inertial range  estimated by Fourier power
spectrum. It indicates that autocorrelation
 function can be used to determine the inertial range.
 Indeed, as we show later, it seems to be  a better inertial range indicator than structure function.

\section{Discussion}
 We  define a cumulative  function
\begin{equation}
\mathcal{Q}(f,\ell,\tau)=\frac{\int_{0}^{f}K(f',\ell,\tau)\upd f'}
{\int_{0}^{+\infty}K(f',\ell,\tau)\upd f'}\label{eq:acfcumulant}
\end{equation}
where
 {\begin{equation} 
K(f,\ell,\tau)=E_{\Delta}(f) \cos(2\pi f \tau)
\end{equation}}
 is the integration kernel of eq.\eqref{eq:acf}. It measures the contribution of the frequency from 0 to $f$ at given scale $\ell$ and time delay $\tau$.
We are particularly concerned by the case $\tau=\ell$.
To avoid the effects of the measurement noise, see fig.\figref{fig:compsp}, we only consider here the   { spanwise (transverse)}  velocity.
We show the estimated $\mathcal{Q}$   in fig.\figref{fig:cumulant}
for two scales  {$\ell/T_s=20$} ($\ocircle$) and  {$\ell/T_s=100$} ($\triangle$)   in the inertial range,
where the  {vertical} solid line illustrates the location of the  corresponding  {time} scale
in spectral space.  {In these experimental curves, the kernel $K$
given in eq.\eqref{eq:acfcumulant} is computed using the experimental spectrum $E_v(f)$.} The corresponding inertial range is denoted by IR. We also show the numerical solution of eq.\eqref{eq:acfcumulant} with $\ell=1$ as inset {, which is estimated by taking a pure power law $E_v(f)\sim f^{-\beta}$ in eq.\eqref{eq:acfcumulant}.} We notice that both curves  cross the line $\mathcal{Q}=0$. We denote  $f_o$ such as $\mathcal{Q}(f_{o})=0$. It has an advantage that the contribution  from large scale $\ell>1/f_o$  is canceled by itself.
Graphically, in the  inertial range, the distance
between $f_o$ and the corresponding scale $\ell$ is less than 0.3 decade.
The numerical solution indicates that this distance is about 0.3 decade.
 We then separate the contribution into a large scale part and a small scale
part. We denote the contribution from the  large scale part as
  {$\mathcal{Q}_1(f)=\mathcal{Q}(1/\ell,\ell,\ell)$}.
The experimental result is shown in fig.\figref{fig:Q1} for both  {streamwise (longitudinal)}
and  {spanwise (transverse)} velocity components. The mean contribution  from large scale is found graphically to be 0.64. It is significantly
larger than 0.5, the value indicated by the numerical solution.  It means  that the autocorrelation function  is influenced more
 by the large scale than by the small scale.

We now  consider the inertial range provided by different methods.
We replot the corresponding compensated spectra estimated  {directly} by Fourier power spectrum
(solid line),  the second order structure function ($\square$), the autocorrelation
function ($\ocircle$) and the Hilbert spectral analysis ($\triangle$) \cite{Huang2008EPL}  in fig.\figref{fig:CP} for  {streamwise (longitudinal)}velocity.
  {For comparison convenience,  both the structure function and the autocorrelation function are converted from physical  space into
spectral space by taking  $f=1/\ell$.}
  For display
convenience, these curves are vertically shifted. Graphically, except the structure
function, the other lines demonstrate a clear plateau. As we have pointed above, the
autocorrelation function is a better indicator of the inertial range than structure
function. We also notice that the inertial range provided by the Hilbert methodology is slightly different from the Fourier spectrum.  This may come from the fact  that the former methodology has a very local ability both in physical and spectral domain \cite{Huang2008EPL,Huang2008TSI}, thus the large scale effect should be constrained. However, the Fourier analysis requires the stationary of the data, which is obviously not satisfied by the turbulence data.
 The result we  present here  can also be linked  with  intermittency property of turbulence: we will present this in future work.

\section{Conclusion}
In this work, we  {considered} the autocorrelation function of the velocity increment  {$\Delta u_{\ell}(t)$ time series, where $\ell$
is a time scale}. Taking  statistical stationary assumption, we propose an
 analytical model of the autocorrelation function.  With this model, we  {proved} analytically that the
location of the minimum autocorrelation function is exactly equal to the separation  {time} scale $\ell$ when the scaling of
the power spectrum of the original variable belongs to the range $0<\beta<2$.
 {In fact, this property was found experimentally to be valid outside
the scaling range, but our demonstration here concerns only the scaling range. }  This
model also suggests a power law expression for the minimum autocorrelation $\Gamma_{o}(\ell)$. Considering
the cumulative integration of  the autocorrelation function{, it was shown that the autocorrelation function is influenced
more by the large scale part}. Finally we argue that the autocorrelation function is a better
indicator of the inertial range  than second order structure function.
 {These results have been illustrated using fully developed turbulence
data; however,  they are of more general validity since we only assumed that
the considered time series is stationary and possesses scaling statistics. }

\acknowledgments This work is supported in part  by the National
Natural Science Foundation of China (No.10772110)
and the Innovation Foundation of Shanghai University. Y.H. is
financed in part by a Ph.D. grant from the French Ministry of
Foreign Affairs. We thank Nicolas Perp\`ete for useful discussion.
Experimental data have been measured in the Johns Hopkins
University's Corrsin wind tunnel and are available for download at
C. Meneveau's web page:
\href{http://www.me.jhu.edu/~meneveau/datasets.html}
{http://www.me.jhu.edu/\~{}meneveau/datasets.html}.

 \bibliographystyle{eplbib}

\end{document}